\documentclass{PoS}

\usepackage{longtable}
\makeatletter
\renewcommand\speaker[1]{\if@speaker\global\@dblspeaktrue\fi
                        \global\@speakertrue
                        \global\setbox\@firstaubox
                        \hbox{\let\thanks\@gobble
                                \let\footnote\@gobble\small 
                                \rm #1}%
                        #1\
                        }%

\newcommand{\PoSlink}[1]{\href{http://pos.sissa.it/cgi-bin/reader/frame_start.cgi?role=reader&action=list.cgi?confid=14}{#1}}

\PoS{PoS(BDMH2004)102}

\title{Baryons in Dark Matter Halos}

\ShortTitle{Baryons in Dark Matter Halos}

\author{Ralph J\"urgen Dettmar\\
Bochum University\\
Bochum, Germany\\
E-mail: \email{dettmar@astro.ruhr-uni-bochum.de}}

\author{Uli Klein\\
Bonn University\\
Bonn, Germany\\
E-mail: \email{uklein@astro.uni-bonn.de}}

\author{\speaker{Paolo Salucci}\\
        SISSA - International School for Advanced Studies\\
	Trieste, Italy\\
        E-mail: \email{salucci@sissa.it}}

\abstract{}

\FullConference{Baryons in Dark Matter Halos\\
               5-9 October 2004\\
               Novigrad, Croatia}

\begin{document}

\global\setbox\@firstaubox\hbox{R.J.~Dettmar, U.~Klein, P.~Salucci}

\makeatother

\section*{Preface}

The existence of Dark Matter (DM) has been long  known. The  first cognitions came  from motions of galaxies in clusters and by the kinematics of individual galaxies  and were followed by systematic investigations, primarily via galaxy rotation curves. Since the mid 90's, observations can be confronted with models defined in specific galaxy formation scenarios, in particular with the output of  simulations performed in the framework of $\Lambda$Cold Dark Matter (CDM). The great success of these models is that they nicely reproduce the large-scale structure, while they - maybe not surprisingly - seem to fail to be equally successful in describing the evolution of the universe on smaller, i.e. cluster and galaxy scales.

These developments were parallelled by breathtaking advancements in cosmology. Since the precision measurement of the spectrum of the cosmic microwave background (CMB) with COBE, subsequent experiments devoted to the CMB anisotropy (Boomerang, WMAP) have led to what is called 'precision cosmology'. This implies that we are in the position of validating (numerical) models to a high degree. At the same time, we are witnessing amazing developments in observational astronomy, which allow to explore the universe back into the epoch of re-ionization, thereby subjecting models to further critical and crucial tests, the last steps expected to be taken in the near future.
All of this does not mean that we may consider most of the riddles solved. In fact, it must be a worry to any astrophysicist that both, DM and Dark Energy remain nothing but hypotheses as long as no particle has been  detected in lab experiments yet.
Nearly ten years of critical validation of CDM models have, alongside with  a lot of success, resulted in what has been coined the "CDM crisis", i.e. the failure of theory to explain certain observational evidences such as    the mass spectrum of satellites around  galaxies, and the (partial) absence of cusps in the dense inner part of galaxy halos.
Another outcome is the realization  that   the  hierarchical picture of  the evolution of  dark matter halos does not imply a similar hierarchy in the baryonic cores  they grow.

The above rationale has been the driving force of the  conference Baryons in Dark Matter Halos,  organized by the Bochum /Bonn graduate research school "Galaxy Groups as Laboratories of Baryonic and Dark Matter" and SISSA,  taken place in Novigrad/Cittanova  5-9 Oct  2004. This event has brought together experts from the whole world  in the relevant  fields, with the aim to make a critical assessment of what has been achieved and to identify the problems that we are faced with and  the future directions of research leading to further progress in our understanding of structure formation and evolution of galaxies.

The Proceedings of the Conference are  published on-line by  PoS (http://pos.sissa.it) which  is a new versatile, online proceedings and lecture notes publication service launched by SISSA, Trieste.  
\bigskip

{\bf R.J. Dettmar, U.  Klein, P. Salucci}

\newpage

\section*{Table of contents}

\renewcommand\arraystretch{1.1}
\noindent\begin{longtable}{p{9em}p{20em}l}\hline
\multicolumn{3}{c}{{\large\sc\bfseries Invited talks}}\\\hline
\raggedright\bf Luigi Danese  & \raggedright\emph{A physical model for formation and evolution of QSOs and of their spheroidal hosts} & \PoSlink{BDMH2004/003}
\\
\raggedright\bf Peter Schuecker  & \raggedright\emph{Present and future applications of galaxy clusters in cosmology} & \PoSlink{BDMH2004/007}
\\
\raggedright\bf Sabine Schindler  & \raggedright\emph{Interaction of galaxies with the intra-cluster me\-dium and ICM metal enrichment} & \href{http://pos.sissa.it/archive/conferences/014/023/schindler.pdf/}{BDMH2004/023}
\\
\raggedright\bf Magda Arnaboldi  & \raggedright\emph{Diffuse light in clusters of galaxies} & \PoSlink{BDMH2004/026}
\\
\raggedright\bf Angela Iovino  & \raggedright\emph{Groups of galaxies} & \PoSlink{BDMH2004/030}
\\
\raggedright\bf Bernd Vollmer  & \raggedright\emph{Galaxy evolution in the Virgo cluster} & \PoSlink{BDMH2004/036}
\\
\raggedright\bf Bianca Poggianti  & \raggedright\emph{Evolution of galaxies in clusters } & \PoSlink{BDMH2004/104}
\\
\raggedright\bf Peter Schneider  & \raggedright\emph{Weak gravitational lensing as a probe of the dark matter distribution} & \PoSlink{BDMH2004/109}
\\
\raggedright\bf Andy Burkert & \raggedright\emph{The structure of cold dark matter halos and the nature of dark matter} & \PoSlink{BDMH2004/105}
\\
\raggedright\bf Frank Van den Bosch  & \raggedright\emph{The galaxy-dark matter connection} & \PoSlink{BDMH2004/041}
\\
\raggedright\bf Rodrigo Ibata   & \raggedright\emph{ The formation of the Milky Way}  & \PoSlink{BDMH2004/106}
\\
\raggedright\bf Reynier Peletier  & \raggedright\emph{The formation of galactic bulges} & \href{http://pos.sissa.it/archive/conferences/014/060/}{BDMH2004/060}
\\
\raggedright\bf Francesca Matteucci  & \raggedright\emph{Chemical evolution of galaxies and galaxy formation mechanisms} & \PoSlink{BDMH2004/072}
\\
\\\hline
\multicolumn{3}{c}{{\large\sc\bfseries Contributed talks}}\\\hline
\raggedright\bf Philipp Richter  & \raggedright\emph{Baryons in the warm-hot intergalactic medium} & \PoSlink{BDMH2004/002}
\\
\raggedright\bf Christian Marinoni  & \raggedright\emph{First-epoch VVDS results: the evolution of the galactic bias up to redshift $z=2$} & \PoSlink{BDMH2004/004}
\\
\raggedright\bf Asmus B\"ohm  & \raggedright\emph{Disk galaxy evolution up to redshift $z=1$} & \PoSlink{BDMH2004/005}
\\
\raggedright\bf Wolfgang Kausch  & \raggedright\emph{Arc statistics with a sample of the most X-ray luminous galaxy clusters} & \PoSlink{BDMH2004/009}
\\
\raggedright\bf Ricardo Genova-Santos  & \raggedright\emph{Searching for the missing baryons with the VSA and WMAP} & \PoSlink{BDMH2004/010}
\\
\raggedright\bf Ester Piedipalumbo  & \raggedright\emph{Some astrophysical implication of gas profiles in a new galaxy clusters model} & \PoSlink{BDMH2004/011}
\\
\raggedright\bf Michela Mapelli  & \raggedright\emph{First stars and extragalactic background light: new constraints to the model through the study of the photon photon absorption} & \PoSlink{BDMH2004/012}
\\
\raggedright\bf Marusa Bradac  & \raggedright\emph{Strong and weak lensing united: the cluster mass distribution of the most X-ray luminous cluster RXJ1347-1145} & \PoSlink{BDMH2004/014}
\\
\raggedright\bf Tim Schrabback  & \raggedright\emph{Cosmic shear with ACS} & \PoSlink{BDMH2004/016}
\\
\raggedright\bf Peter Kalberla  & \raggedright\emph{Baryonic dark matter in the Milky Way} & \PoSlink{BDMH2004/018}
\\
\raggedright\bf Leonidas Dedes  & \raggedright\emph{Spiral structure and the clumpy HI sub-structure of the halo of the Milky Way} & \PoSlink{BDMH2004/019}
\\
\raggedright\bf Liliya L. R. Williams  & \raggedright\emph{The phase-space density distribution of dark matter halos} & \PoSlink{BDMH2004/020}
\\
\raggedright\bf Matteo Viel  & \raggedright\emph{Inferring the dark matter power spectrum from the Lyman-Alpha forest in high-resulotion QSO absorption spectra} & \PoSlink{BDMH2004/021}
\\
\raggedright\bf Wolfgang Kapferer  & \raggedright\emph{Hydrodynamic galaxy cluster simulations: a challenge for physics, parallel computng and visualisation} & \PoSlink{BDMH2004/024}
\\
\raggedright\bf Dominik Rosenbaum  & \raggedright\emph{The enviroment of low surface brightness galaxies} & \PoSlink{BDMH2004/025}
\\
\raggedright\bf Alexis Finoguenov  & \raggedright\emph{XMM-Newton survey of IGM: news for the modified entropy scaling} & \PoSlink{BDMH2004/028}
\\
\raggedright\bf D.J. Pisano  & \raggedright\emph{Low mass dark matter halos in loose groups of galaxies} & \PoSlink{BDMH2004/031}
\\
\raggedright\bf Elvira Krusch  & \raggedright\emph{Investigation of the dwarf galaxy population in Hickson Compact Groups} & \PoSlink{BDMH2004/033}
\\
\raggedright\bf Leonardo Castenada-Co\-lo\-ra\-do  & \raggedright\emph{Kinematics in Hickson compact group 90} & \PoSlink{BDMH2004/034}
\\
\raggedright\bf Lucio Mayer  & \raggedright\emph{Baryons in SPH simulation of structure formation and evolution; approaching the end of the dark era} & \PoSlink{BDMH2004/037}
\\
\raggedright\bf Michael Hilker  & \raggedright\emph{The properties of ultra-compact dwarf galaxies and their possible origin} & \PoSlink{BDMH2004/038}
\\
\raggedright\bf Eva Manthey  & \raggedright\emph{Properties of moderate luminosity mergers} & \PoSlink{BDMH2004/040}
\\
\raggedright\bf Gianfranco Gentile  & \raggedright\emph{Lambda CDM and the dark matter distribution in spirals} & \PoSlink{BDMH2004/042}
\\
\raggedright\bf Gyula J\'ozsa  & \raggedright\emph{Dynamics of warped disk galaxies} & \PoSlink{BDMH2004/043}
\\
\raggedright\bf Alessandro Pizzella  & \raggedright\emph{Low surface brighness galaxies: Vc-s0 relation and halo central density radial profile from stellar kinematics measurements} & \href{http://pos.sissa.it/archive/conferences/014/046/apizzella5.0.pdf}{BDMH2004/046}
\\
\raggedright\bf Isabel Perez Martin  & \raggedright\emph{Dark matter in the inner parts of barred galaxies} & \PoSlink{BDMH2004/047}
\\
\raggedright\bf Masayuki Tanaka  & \raggedright\emph{PISCES: galaxy properties as functions of enviroment and time} & \PoSlink{BDMH2004/048}
\\
\raggedright\bf Enrico Maria Corsini  & \raggedright\emph{The dark matter content of early-type barred galaxies} & \PoSlink{BDMH2004/049}
\\
\raggedright\bf Aaron Dutton  & \raggedright\emph{Scaling relations of spiral galaxies: theory vs.\ observation} & \PoSlink{BDMH2004/050}
\\
\raggedright\bf David Roscoe  & \raggedright\emph{New phenomenological constraints for dark matter models in disks} & \PoSlink{BDMH2004/051}
\\
\raggedright\bf Fabio Fontanot  & \raggedright\emph{High-redshift QSOs in GOODS} & \PoSlink{BDMH2004/053}
\\
\raggedright\bf Giuseppina Battaglia  & \raggedright\emph{Kinematics and metallicity relations for dwarf galaxies in the local group} & \PoSlink{BDMH2004/054}
\\
\raggedright\bf Riccardo Scarpa  & \raggedright\emph{Using globular clusters to test gravity in the weak acceleration regime: NGC 6171} & \PoSlink{BDMH2004/055}
\\
\raggedright\bf Francesca Annibali  & \raggedright\emph{Spectro-photometric predictions of a model for the joint formation of QSOs and spheroids} & \PoSlink{BDMH2004/056}
\\
\raggedright\bf Tobias Kaufmann  & \raggedright\emph{Numerical influences on galaxy formation} & \PoSlink{BDMH2004/057}
\\
\raggedright\bf Yves Revaz  & \raggedright\emph{Bending Instabilities at the origin of persistant warps: a constraint on the dark matter} & \PoSlink{BDMH2004/058}
\\
\raggedright\bf Francesco Shankar  & \raggedright\emph{The baryonic vs.\ dark matter halo mass relationship in galaxies: the effect of the inefficiency of the cosmological star formation} & \PoSlink{BDMH2004/059}
\\
\raggedright\bf Giuseppe Aronica  & \raggedright\emph{Peanut shaped structures in edge-on galaxies} & \PoSlink{BDMH2004/061}
\\
\raggedright\bf Carsten Weidner  & \raggedright\emph{Supernova-rates for different galaxy types} & \PoSlink{BDMH2004/063}
\\
\raggedright\bf Ylva Schuberth  & \raggedright\emph{Kinematics of the outer cluster system of NGC 1399} & \href{http://pos.sissa.it/archive/conferences/014/065/1399.outer.rev.new.pdf}{BDMH2004/065}
\\
\raggedright\bf Leon Koopmans  & \raggedright\emph{Dark-matter and baryons in early-type lens galaxies} & \PoSlink{BDMH2004/066}
\\
\raggedright\bf Nicola Napolitano  & \raggedright\emph{Dark-to-luminous properties of early type galaxies} & \PoSlink{BDMH2004/067}
\\
\raggedright\bf Edo Noordermeer  & \raggedright\emph{Rotation curves and dark matter in early type disk galaxies} & \PoSlink{BDMH2004/068}
\\
\raggedright\bf Lynn Matthews  & \raggedright\emph{Clues on structure and composition of galactic disks from studies of `superthin' spirals} & \PoSlink{BDMH2004/070}
\\
\raggedright\bf Francesco Calura  & \raggedright\emph{Cosmic star formation history: pure luminosity vs.\ number galaxy evolution} & \PoSlink{BDMH2004/073}
\\
\raggedright\bf Wilfried Domainko  & \raggedright\emph{Metal enrichment of the intra-cluster medium: ram-pressure stripping and feedback from intra-cluster supernovae} & \PoSlink{BDMH2004/074}
\\
\raggedright\bf Alessio D. Romeo  & \raggedright\emph{Simulating galaxy clusters: the ICM and the galaxy populations} & \PoSlink{BDMH2004/075}
\\
\raggedright\bf Daniel Pfenniger  & \raggedright\emph{Dark molecular hydrogen} & \PoSlink{BDMH2004/087}
\\
\raggedright\bf Andrea Biviano  & \raggedright\emph{The relative distribution of dark matter and baryons in galaxy clusters} & \PoSlink{BDMH2004/088}
\\
\raggedright\bf Antonio Pipino  & \raggedright\emph{Formation and evolution of massive elliptical galaxies in clusters: a consistent picture from optical and X-ray properties} & \PoSlink{BDMH2004/091}
\\
\raggedright\bf Antti Tamm  & \raggedright\emph{Structure of visual and dark matter components of spiral galaxies at $z \sim 1$} & \PoSlink{BDMH2004/092}
\\
\raggedright\bf Jelte De Jong  & \raggedright\emph{Probing MACHOs in M31} & \PoSlink{BDMH2004/093}
\\
\raggedright\bf Matthias Hoeft  & \raggedright\emph{Galaxy formation in voids} & \PoSlink{BDMH2004/094}
\\
\raggedright\bf Peter Erni  & \raggedright\emph{The damped Lyman alpha absorber toward Q0913+072} & \PoSlink{BDMH2004/096}
\\
\raggedright\bf Patrick Simon  & \raggedright\emph{The galaxy-dark matter bias} & \PoSlink{BDMH2004/097}
\\
\raggedright\bf Anatoli Iyudin  & \raggedright\emph{New gamma-ray probe of the baryonic dark matter} & \PoSlink{BDMH2004/098}
\\
\raggedright\bf Liliya L. R. Williams  & \raggedright\emph{Strong lensing} & \PoSlink{BDMH2004/100}
\\
\raggedright\bf Michael Fellhauer  & \raggedright\emph{How star clusters could survive low star formation efficiencies} & \PoSlink{BDMH2004/101}
\\
\raggedright\bf Pieter Buyle  & \raggedright\emph{Beyond the sphericitiy assumption in dynamical HI models} & \PoSlink{BDMH2004/076}
\\
\raggedright\bf Franz Kenn  & \raggedright\emph{The dark halo in the spiral galaxy NGC 755} & \PoSlink{BDMH2004/077}
\\
\raggedright\bf Ivana Damjanov  & \raggedright\emph{Fuelling star-formation --- the fate of halo baryons?} & \PoSlink{BDMH2004/078}
\\
\raggedright\bf Marco Hetterscheidt  & \raggedright\emph{Searching for clusters using weak lensing} & \PoSlink{BDMH2004/079}
\\
\raggedright\bf Enrichetta Iodice  & \raggedright\emph{High resolution stellar kinematics for NGC4650A: solving the enigma of the flattening of its dark halo} & \PoSlink{BDMH2004/081}
\\
\raggedright\bf Thomas Kronberger  & \raggedright\emph{Dark matter in numerical simulations of galaxy clusters} & \PoSlink{BDMH2004/082}
\\
\raggedright\bf Jairo Mendez Abreu  & \raggedright\emph{Measuring bulge and disk surface brightness in disk galaxies} & \PoSlink{BDMH2004/083}
\\
\raggedright\bf Irina Yegorova  & \raggedright\emph{The tully Fisher relation of spiral galaxies} & \PoSlink{BDMH2004/084}
\\
\raggedright\bf Chiara Tonini  & \raggedright\emph{Mass modelling from rotation curves} & \PoSlink{BDMH2004/089}
\\
\raggedright\bf Oliver Czoske  & \raggedright\emph{A wide-field spectroscopic survey of Abell 1689 and Abell 1835 with VIMOS} & \PoSlink{BDMH2004/099}
\\
\end{longtable}

\end{document}